\documentclass[A4,prl, twocolumn, superscriptaddress ,preprintnumbers,amsmath,amssymb,floatfix]{revtex4}
\usepackage{color}
\usepackage{graphicx}% Include figure files
\usepackage{dcolumn}% Align table columns on decimal point
\usepackage{bm}% bold matheq:
\usepackage[mathscr]{eucal}
\usepackage[nooneline]{subfigure}

\newcommand{\ket}[1]{\vert#1\rangle}

\begin{document}

\title{Response to Comment on ``Continuous quantum measurement: inelastic tunnelling suppresses current oscillations''}

\author{T.M. Stace}%\email{tms29@cam.ac.uk}
\affiliation{Cavendish Laboratory and DAMTP, University of
Cambridge, Cambridge, UK}
\author{S. D. Barrett}%\email{sean.barrett@hp.com}
\affiliation{Hewlett Packard Laboratories,  Filton Road, Stoke
Gifford Bristol, BS34 8QZ, UK}


\textbf{Stace and Barrett reply:} Our recent work
\cite{stace922004} considered a system consisting of a charge
qubit coupled to a point contact (PC) charge detector in the
relatively unexplored parameter regime, $\Gamma_d\ll\phi$, where
$\Gamma_d$ is the measurement-induced dephasing rate of the
detector upon the qubit, and $\phi$ is the qubit energy splitting.
Describing the \emph{conditional} evolution of the system in this
parameter regime is particularly interesting in light of recent
experiments \cite{gar03,elz03,dicarlo:226801}, in which
phenomena such as partial localization of an electron in the
energy eigenstates of a double well system have been observed.
Previous theoretical analyses of the conditional dynamics of this
system \cite{kor01b,goa01a} have been restricted to the
limit of large detector voltage bias, $e V \gg \phi$, where, in
the steady state, localization does not occur. On the other hand,
our analysis is valid for \emph{arbitrary} detector bias, since it
properly takes account of qubit relaxation processes due to
inelastic tunnelling in the detector.

In \cite{stace922004} we made a number of predictions about the
unconditional and conditional dynamics in both the high- ($e V >
\phi$) and low-bias ($e V < \phi$) regimes. In their comment, Averin and Korotkov (AK)
take issue with just one of these predictions: our claim that
coherent oscillations are absent in the detector output (which is
contrary to previous claims). AK presume that this discrepancy stems
from an assumption ``...\ that the qubit interaction with
the PC detector suppresses quantum interference between qubit
energy eigenstates.''.  Furthermore, they assert that the form of our expression for the current (Eq.\
(9) of \cite{stace922004}), containing three
`jump' operators, follows by assumption. 

In fact, we made no such assumptions.  Rather Eq.\ (9) follows naturally from Eq.\ (5), which is in turn rigourously derived from a microscopic model, as we shall now review.  We began by deriving an unconditional master equation (UME).  In
common with previous analyses of this system \cite{goa01a}, we make
the Born-Markov approximation, which assumes factorised initial
conditions and rapid relaxation of the environment (PC leads).
Following this, we make a rotating wave approximation (RWA).  The
RWA represents an equation for the the lowest order term in a
power series expansion of the density matrix, in the perturbative
parameter $\Gamma_d/\phi$. This procedure results in a Markovian
UME with three Lindblad super-operators.

We proceed to `unravel' this UME to produce a conditional master
equation (CME), capable of describing the dynamics of the qubit
conditional upon the stochastic measurement results. The CME must
be consistent with the UME, so it follows that the CME has three
`jump' operators, arising from the three Lindblad super-operators.
Thus the number of jump processes is \emph{not} an arbitrary
assumption, as claimed in the comment, but is a necessary
consequence of the UME in the limit $\Gamma_d\ll\phi$, for any
finite value of the ratio $e V/\phi$. In \cite{stace922004} the
particular form of our jump operators is determined by physical
considerations, such as energy conservation, but they can also be
derived directly from an explicit model of the measurement process
(see Ref. \cite{sta03c}).

In the low bias regime, $e V<\phi$, the UME predicts that the
qubit relaxes to the (pure) ground state, $\ket{g}$. Since
$\ket{g}$ is stationary, there should be no oscillatory signal in
the PC current, and no peaks in $S_{\mathrm{lb}}(\omega)$ should
be seen at $\omega=\phi$, in agreement with \cite{shn02}.

Although the specific objections raised by AK are unfounded, we note that in the high-bias regime, $e V>\phi$, there may be a problem in interpreting our power spectra at high frequencies. 
In making the RWA to arrive at the UME, we have ignored fast dynamics on time scales $\sim\phi^{-1}$.  In \cite{sta03c} this temporal `coarse-graining' is made evident by deriving the jump operators using an explicit current measurement model in which PC tunnelling events are counted.
Therefore, although our expression for $S_{\mathrm{hb}}(\omega)$
is correct for $\omega\ll\phi$, it may not apply at frequencies comparable to $\phi$, in the high-bias regime.

In summary, we now believe that our assertion in
\cite{stace922004}, that coherent oscillations in the detector
output are suppressed, may only be justified in the low-bias regime, $e V < \phi$. In the high-bias regime, $e V > \phi$, we are unable to make firm
predictions about such high-frequency oscillations, since our Markovian
description only strictly applies for timescales longer than $\phi^{-1}$.
The remainder of our conclusions are valid. Furthermore, our approach provides an accurate description of continuous measurement in experimentally accessible parameter regimes, and will serve as a basis for future theoretical work.

We thank Howard Wiseman, Kurt Jacobs, Neil Oxtoby, Gerard Milburn, Hsi-Sheng Goan and Dmitri Averin for useful conversations.

\renewcommand{\url}[1]{}
\newcommand{\urlprefix}[1]{#1}
%\bibliography{tomsbib}
%These suppress URL information from appearing in the bibliography

%\input{ResponseToCommentPRLLowBias.bbl}

\end{document}